\documentclass[preprint,bibnotes,floats,aps]{revtex4}
\usepackage{times} 
\usepackage{graphicx}
\graphicspath{{.},{$HOME/Obj/Rundrop/Ps/},{$HOME/Obj/Sitdrop/Tex/Ps/}}
\begin{document}

\bibliographystyle{$HOME/Literatur/Bibdir/revtex}

\title{Dynamical Model for Chemically Driven Running Droplets}
\author{Uwe Thiele}
\email{thiele@mpipks-dresden.mpg.de, http:\\www.uwethiele.de}
\author{Karin John}
\email{john@mpipks-dresden.mpg.de}
\author{Markus B{\"a}r}
\email{baer@mpipks-dresden.mpg.de}
\affiliation{Max-Planck-Institut
f\"ur Physik komplexer Systeme, N{\"o}thnitzer Str.\ 38, D-01187 Dresden, Germany}
\begin{abstract}
We propose coupled evolution equations for the thickness of a liquid film 
and the density of an adsorbate layer on a partially wetting solid substrate. 
Therein, running droplets are studied assuming 
a chemical reaction underneath the droplets that induces a
wettability gradient on the substrate and provides the driving 
force for droplet motion. 
Two different regimes for moving droplets -- reaction-limited and saturated
regime -- are described. They correspond 
to increasing and decreasing velocities with increasing reaction
rates and droplet sizes, respectively. The existence of the two regimes offers a 
natural explanation of  prior experimental observations. 
\end{abstract}
%
\pacs{
68.15.+e, 
47.20.Ky  
68.08.-p  
68.43.-h  
}

%
\maketitle

%
%
It is well known that a drop of liquid can start to move if it is exposed
to an external gradient. 
For instance,
a drop of liquid immersed in another liquid that is placed in a temperature 
gradient will move towards the higher temperature due to Marangoni forces caused by
surface tension gradients \cite{Vela98}.
Alternatively, a temperature gradient \cite{Broc89}
or a wettability gradient induced by a chemical grading 
of the substrate \cite{Raph88,ChWh92} cause drop motion. 

Drops may also move in initially homogeneous surroundings. 
This is possible because such 
{\it active drops} change their surrounding and 
produce an internal gradient that drives their motion.
One example are drops immersed in another liquid  that contain 
a soluble surfactant undergoing an isothermal chemical reaction at the
surface of the drop \cite{RRV94,MiMe97}.
Recent experiments found a different sort of chemically driven 
running droplets on solid substrates. 
There, the  driving wettability gradient is produced 
by a chemical reaction at the substrate underneath the drop 
\cite{BBM94,DoOn95,LKL02}. 

\begin{figure}[tbh]
\includegraphics[width=0.9\hsize]{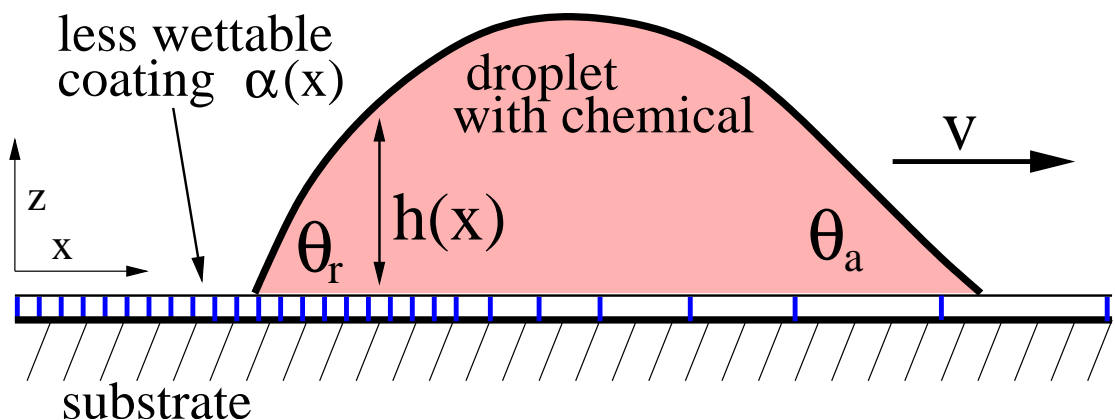}
\caption{Sketch of a right moving droplet driven by a chemical gradient.
}
\label{sketch}
\end{figure}

In these experiments, a small droplet of solvent is put on a
partially wettable substrate. 
A chemical dissolved in the
droplet starts to react with the substrate resulting in the 
deposition of a less wettable coating. 
The substrate below becomes 
less wettable than the substrate outside the droplet.
Eventually, the symmetry is broken by fluctuations 
and the drop starts to move away from the 'bad spot', 
thereby changing the substrate and leaving a less wettable trail behind 
(see Fig.\,\ref{sketch}).
Similar phenomena can be seen in reactive (de)wetting \cite{ZWT98}, in 
camphor boats \cite{Haya02}, and
in the migration of reactive islands in alloying \cite{SBH00}. 

From a simple theoretical argument \cite{Brde95},  
an implicit equation for the velocity $v$ of the droplet was derived 
\cite{Brde95}: $v = C \tan \theta^{*} ( 1 - \exp(-r L/v ))$, where 
$r$ is the reaction rate, $L$ the droplet's length and $C$ a constant.  
The dynamic contact angle $\theta^{*}$ is assumed to be same at
the advancing and receding ends of the droplet, i.e.\ the droplet 
profile is approximated by a spherical cap. 
$\theta^{*}$ is given by
$\cos \theta^{*} = (\cos \theta_e^a + \cos \theta_e^r)/2$, 
where  the static contact angles at the advancing edge $\theta_e^a$ and  
at the receding edge $\theta_e^r > \theta_e^a$ are different due to the chemical
gradient. 
The expression for the velocity is found assuming a first order
reaction on the substrate yielding chemical concentrations $\alpha_a = 0$ 
and $\alpha_r =  1 - \exp(-r L/v) $ at the respective ends of the 
droplet. 
The expression for the velocity predicts a monotone increase of the
droplet velocity with the droplet length $L$ and the reaction rate
$r$, in line with early experimental observations \cite{DoOn95}.  
Recent experiments \cite{LKL02} show the opposite trend; the velocity 
decreases with increasing drop sizes and reaction rate. 

In this Letter we propose and analyze a dynamical model for self-propelled 
running droplets.
It consists of coupled evolution equations describing the
interdependent spatiotemporal dynamics of the  film thickness $h$ 
and a chemical concentration $\alpha$ that changes the substrate
wettability.
This model is capable of reproducing 
both experimentally found regimes. 
In particular, we identify two distinct regimes  of running drops
separated by a maximum of the velocity.
For small droplet size or reaction rate, the chemical gradient in 
the drop is limited by the progress of reaction. 
In contrast for large droplets and fast reaction, the chemical
concentration at the receding end saturates at a maximum value. 
Below, we show that the velocity of reaction-limited droplets 
increases, while the velocity of saturated
droplets decreases with increasing reaction rate and drop size.
In addition, we find that the dynamic contact angles at the advancing
and receding edges of the droplets differ substantially.  
If we take into account fast diffusion of the chemical substance, 
the droplet velocity in the saturated regime decreases much faster and
only stationary drops are found at large reaction rates. 
The velocity of the droplets may increase \cite{DoOn95} or decrease 
\cite{LKL02} with their volume.

A liquid layer on a smooth solid substrate is modelled by an
evolution equation for the film thickness profile $h(\vec{x},t)$ 
derived from the Navier-Stokes equations using long-wave approximation
\cite{ODB97}
\begin{equation}
\partial_t\,h\,=\,-\nabla\,\cdot \left[\frac{h^3}{3\eta}\, 
\nabla\,(\gamma\Delta h + \Pi(h)) \right]
\label{film}
\end{equation}
with $\gamma$ and $\eta$ being the surface tension and viscosity
of the liquid, respectively. 
The disjoining pressure $\Pi(h)$
accounts for the wetting properties of the substrate \cite{Isra92}.
We use the form $\Pi(h)=2S_a d_0^2/h^3 + (S_p/l)\,\exp[(d_0-h)/l]$ 
\cite{Shar93,TNPV02}, where $S_a$ and $S_p$ are the apolar and polar components
of the total spreading coefficient $S=S_a+S_p$, respectively, 
$d_0=0.158\,nm$ is the Born repulsion length and $l$ is a correlation 
length \cite{Shar93}. 
We choose $S_a>0$ and $S_p<0$ 
thereby combining a stabilizing long-range van der Waals
and a destabilizing short-range polar 
interaction. 
The latter contains the influence of the coating and
crucially influences the static contact angle \cite{Shar93}.  
Such a model allows
for solutions representing static droplets with a finite mesoscopic
equilibrium contact angle sitting on an ultrathin precursor film.

To account for the varying wettability caused by the chemical 
reaction 
we let the polar part of the spreading coefficient $S_p$
depend linearly on the coating density, i.e.\ on the 
concentration $\alpha(\vec{x},t)$ of a chemical species adsorbed at the substrate:
\begin{equation}
S_p\,=\,S_p^0\,(1+\frac{\alpha}{g'}) < 0.
\label{spread}
\end{equation}
The equilibrium contact angle $\theta_e$ is given by
$\cos\theta_e=S/\gamma+1$  \cite{Shar93}. 
This implies that $\theta_e$  increases with  $\alpha$, i.e.\ the coated
substrate is less wettable. 
The constant $S_p^0/g'$  defines the magnitude of the wettability
gradient. 
Note, that Eq.\,(\ref{spread}) corresponds to the linear relation
between $\cos \theta_e$  and $\alpha$  assumed in \cite{DoOn95,LKL02,Brde95}.

The evolution of the chemical concentration on the substrate  is modelled
by a reaction-diffusion equation for $\alpha(\vec{x})$

\begin{equation}
\partial_t\,\alpha\,=\,R(h,\alpha) \,+\, d'\,\Delta\alpha,
\label{rd}
\end{equation}
where the function 
$R(h,\alpha)$ describes the reaction that changes
the wettability of the substrate and the second term
allows for a (usually small) diffusion of the chemical species along the 
substrate.  
The main results, however,  are obtained without
diffusion. 
As reaction term we choose
\begin{equation}
R(h,\alpha)\,=\,r' \Theta(h-h_0) \,\left(1-\frac{\alpha}{\alpha_{\text{max}}}\right).
\label{reac}
\end{equation}
The time scale of the reaction is defined by the rate constant $r'$. 
It is assumed that the reaction at the substrate occurs only 
underneath the droplet; this is modelled by the step function $\Theta(h-h_0)$. 
The value of $h_0$ is chosen slightly
larger than the thickness of the precursor film. 
The chemical concentration of the coating saturates at a value
$\alpha_{\text{max}}$,
because the reaction is assumed to be fast enough to produce a
complete adsorption layer.
Changes in details of the reaction term like the replacement of
the step function by an explicit proportionality to $h$
do not affect our results qualitatively. 
The physical motivation behind the actual choice is a fast
equilibration of the reactant concentration within the moving droplet,
caused by diffusion and convective motion within the droplet.

\begin{figure}[tbh]
\includegraphics[width=0.9\hsize]{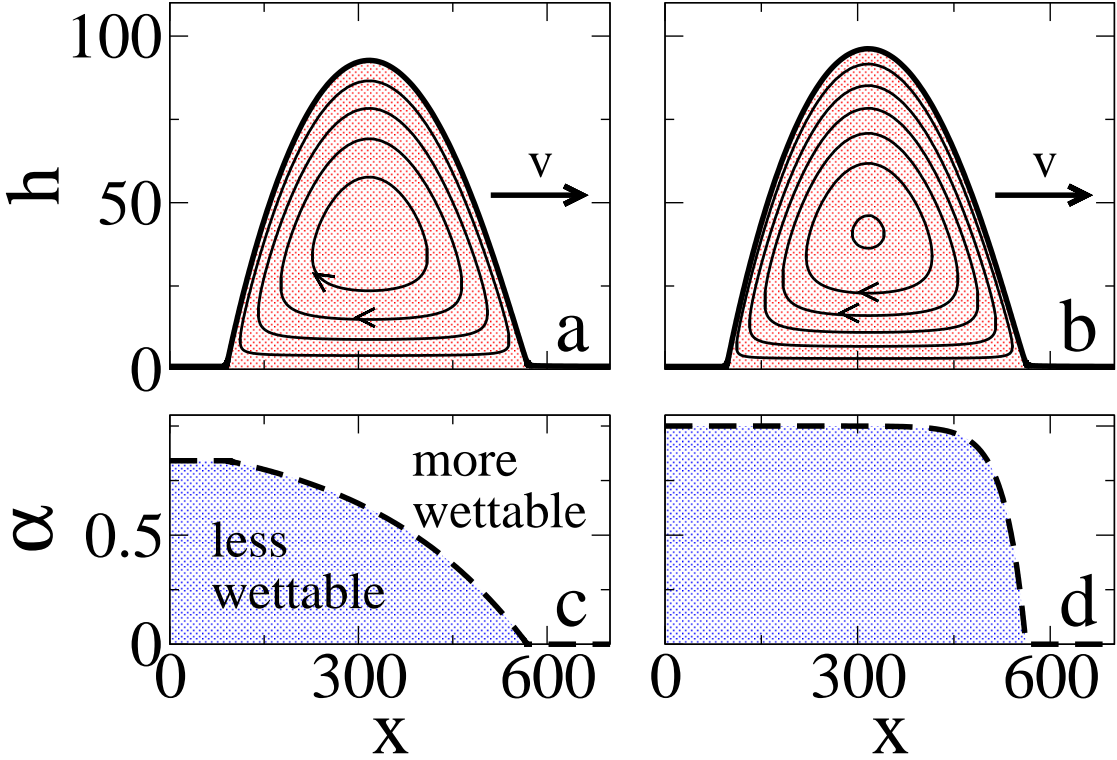}
\caption{Profiles of the droplet $h(x,t)$  ((a) and (b)) and 
concentration $\alpha$ of coating ((c) and (d)) in the two qualitatively 
different running droplet regimes. 
(a) and (c) Reaction-limited regime
at $r=0.0001$ moving with velocity $v=0.026$. 
(b) and (d) Saturated regime at $r=0.001$  with $v=0.032$. 
The streamlines plotted in (a) and (b) 
indicate the convective motion inside the droplet in the comoving frame \cite{boundary}. 
The remaining parameters are $g=1.0, d=0.0, b=0.5$ and the droplet volume is
30000.}
\label{profiles}
\end{figure}

We rewrite 
Eqs.\,(\ref{film}) to (\ref{reac})
by introducing scales $3\gamma\eta/l\kappa^2$, $\sqrt{l\gamma/\kappa}$,
$\alpha_{\text{max}}$ and $l$ for $t$, $\vec{x}$, $\alpha$ and $h$,
respectively.
Then, $\kappa=(|S_p|/l)\exp(d_0/l)$ is the dimensionless spreading 
coefficient. 
Defining the dimensionless reaction rate 
$r=3r'\gamma\eta/l\kappa^2$, 
diffusion constant $d=3d'\eta/\kappa l^2$,
wettability gradient $g=g'/\alpha_{\text{max}}$ and 
ratio of the effective molecular interactions 
$b=2S_ad_0^2/l^3\kappa$, we obtain from Eqs.\,(\ref{film}) to (\ref{reac}) 
the dimensionless coupled evolution equations for the thickness profile $h$
and the substrate coverage $\alpha$
\begin{eqnarray}
\partial_t\,h\,&=&\,-\nabla\,\cdot\left\{h^3\, 
\nabla\,\left[\Delta h
+  \frac{b}{h^3} - 
\left(1+\frac{\alpha}{g}\right)\,e^{-h} \right]\right\} \label{sys1}\\
\partial_t\,\alpha\,&=&\,r \Theta(h-h_0)\,(1-\alpha)
\,+\,d\,\Delta\alpha
\label{sys2}
\end{eqnarray}
In the one-dimensional version of Eqs.\,(\ref{sys1},\ref{sys2}), we calculate
by use of continuation techniques \cite{AUTO97} two-dimensional running
droplets moving with constant speed.
This is achieved by switching to the comoving frame $x-vt$ and
imposing appropriate boundary conditions \cite{boundary}.  

Fig.\,\ref{profiles} shows the constant profiles of such moving droplets ((a) and (b)) 
together with the corresponding concentration profiles for $\alpha$ ((c) and (d)) 
for two qualitatively different regimes 
(prominently visible in the $\alpha$ profiles). 
In ((a) and (b)) the streamlines in the comoving frame indicate the
convective motion in the droplets.

In Figs.\,\ref{profiles}\,(a,c)  the limiting factor for the motion
is the small reaction rate. 
The reaction time is long compared to the 
viscous time scale, 
i.e.\ the droplet passes too fast for the coverage $\alpha$
to approach its saturation value. 
In the case $d=0$, we can estimate the value of $\alpha$ at
the receding end to be $\alpha_r = 1 - \exp ( - r L/ v)$ where $L$ is 
the length of the droplet. 
We call this regime reaction-limited.
In Figs.\,\ref{profiles}\,(b,d) 
the reaction at the receding edge has reached its 
saturating value, i.e.\ $\alpha_r \approx 1$ and $r L/v >> 1$. 
After the droplet has passed, the concentration $\alpha$ stays at the 
saturation level. 
We call this regime saturated. 
One can not exactly predict where the transition
between the regimes occurs, because the velocity
of the droplet and the driving wettability gradient determine each other
dynamically.
We found that in both regimes  the advancing contact angle is substantially smaller 
than the receding one. 
This contradicts the assumption of identical dynamic contact angles
in the approximate theory \cite{Brde95}.

\begin{figure}[h]
\begin{center}
\includegraphics[width=0.8\hsize]{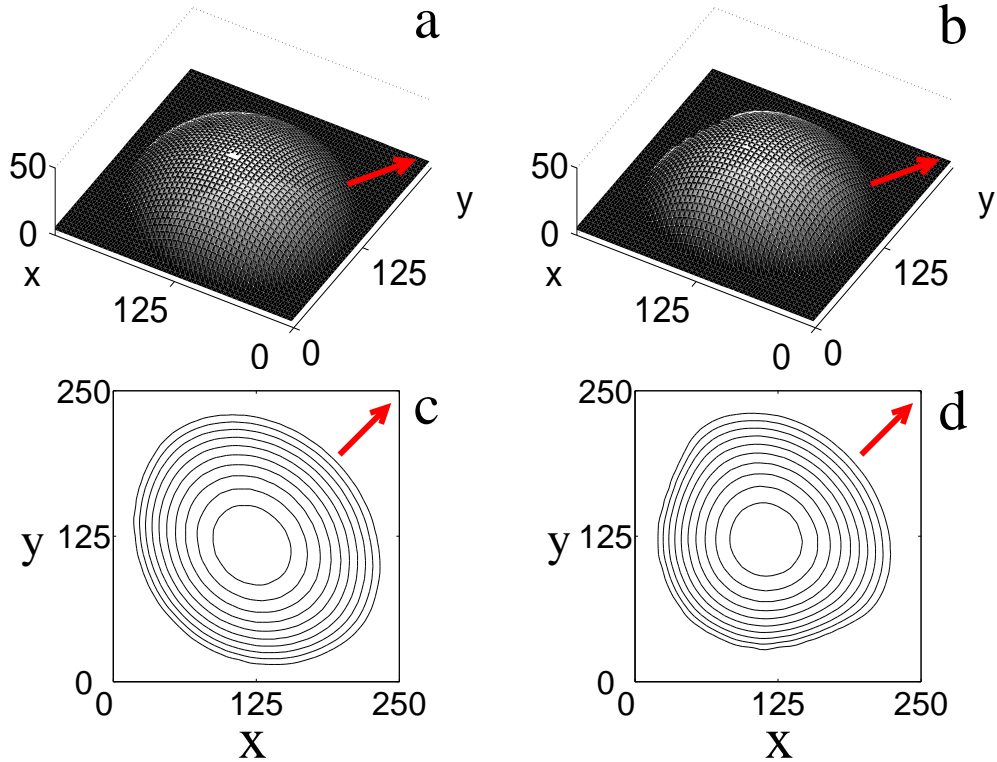}
\end{center}
\caption{
Moving drops in three dimensions. 
Shown are the film thickness profiles
in the (a) reaction-limited and (b) saturated regime.
(c) and (d) represent the corresponding contour lines.
The velocity converged to $v\approx0.018$ and $v\approx0.044$, respectively.
The direction of movement is indicated by arrows,
the parameters correspond to Figs.\,\ref{profiles}\,(a,c),
respectively, and the droplet volume is $7.5\times10^5$.
Only a part of the computational domain size of $500\times500$ is shown.}
\label{2d}
\end{figure}

\begin{figure}[tbh]
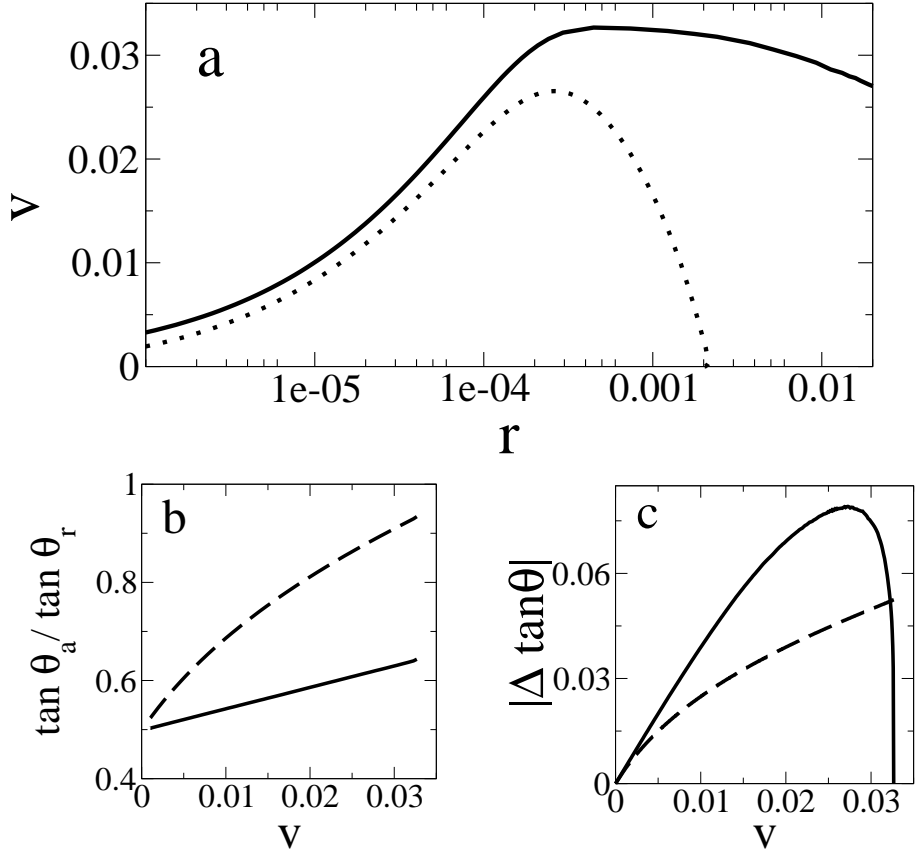

\begin{center}
\includegraphics[width=0.7\hsize]{VvsR1_c.eps}\\[1ex]
\hspace{0.05\hsize}\includegraphics[width=0.33\hsize]{ContAngvsV_c.eps}\hspace{0.05\hsize}
\includegraphics[width=0.33\hsize]{differencescontang_d0001b.eps}
\end{center}
\caption{Characterization of running droplets stationary in a 
comoved frame. Shown are (a) the velocity $v$ in dependence of the reaction rate $r$
(logarithmic scale) without ($d=0$, solid line) and with ($d=1.0$, dotted line) diffusion.;
(b) the tangents of the advancing $\theta_a$ (solid line) and receding
$\theta_r$ (dashed line) dynamic contact
angles in the reaction-limited regime; (c) the absolut value of the difference
of $\tan\theta_a$ (solid line) 
and $\tan\theta_r$  (dashed line) and the tangents of the static contact angle
$\theta_e(\alpha)$ at the corresponding $\alpha$ values. 
Also in (c) only the reaction-limited regime
$r<r_m$ is shown where always $\theta_a>\theta_e(\alpha_a)$ and 
$\theta_r<\theta_e(\alpha_r)$. The remaining parameters are as in Fig.\,\ref{profiles}.
}
\label{vel}
\end{figure}

We have tested that the above  model also allows for stationary moving droplets in 
three dimensions (Fig.\,\ref{2d}) by integrating Eqs.\,(\ref{sys1}) and 
(\ref{sys2}) in time starting from adequate initial conditions \cite{note2}.
After a transient the droplets approach constant shape and velocity.
We show resulting droplets for the reaction-limited (Figs.\,\ref{2d}\,(a,c))
and the saturated (Figs.\,\ref{2d}\,(b,d)) regime. 
Both drops have oval shape with an asymmetry between advancing and receding part 
that is stronger in the 
saturation-limited regime due to the larger wettability gradient.

A closer analysis of the one-dimensional version of
Eqs.\,(\ref{sys1},\ref{sys2}) gives insight
into the properties of the running droplets and their existence and stability
conditions. 
Fig.\,\ref{vel}\,(a) shows the droplet velocity
in dependence of the reaction rate $r$ without and with diffusion.
The dependence of the velocity on
reaction rate $r$ for the running droplets is non-monotonic.
There exists a value $r_m$ where the velocity is maximal. 
The case $r<r_m$ [$r>r_m$]  defines the reaction-limited 
[saturated] regime. 

The velocity increase for $r<r_m$ is easily explained
by the increasing value of $\alpha_r$ at the back of the drop  caused
by the faster reaction. 
The decreasing velocity for $r>r_m$ originates from a slow increase
of $\alpha$ in the contact region at the front of the drop,
while $\alpha$ at the back remains constant at its saturation value 
($\alpha_r \approx 1)$. 
We have also studied the case where diffusion of $\alpha$ plays is sizable.
There,
the rapidly produced $\alpha$ diffuses to the substrate in front of
the droplet and diminishes the driving gradient between
front and back of the drop. 
The velocity curve is similar for small $r$ but decreases even faster beyond
the maximum and drops to zero at large enough $r$; 
there, no running droplets are found anymore. 
The latter phenomena is reminiscent to observations of moving spots in
reaction-diffusion systems of activator-inhibitor-type \cite{KrMi94,OBSP98,Boed03}.
Therein, a transition from moving to stationary droplets (drift
pitchfork bifurcation) occurs when the inhibitor is produced to fast. 

\begin{figure}[h]
\begin{center}
\includegraphics[width=0.8\hsize]{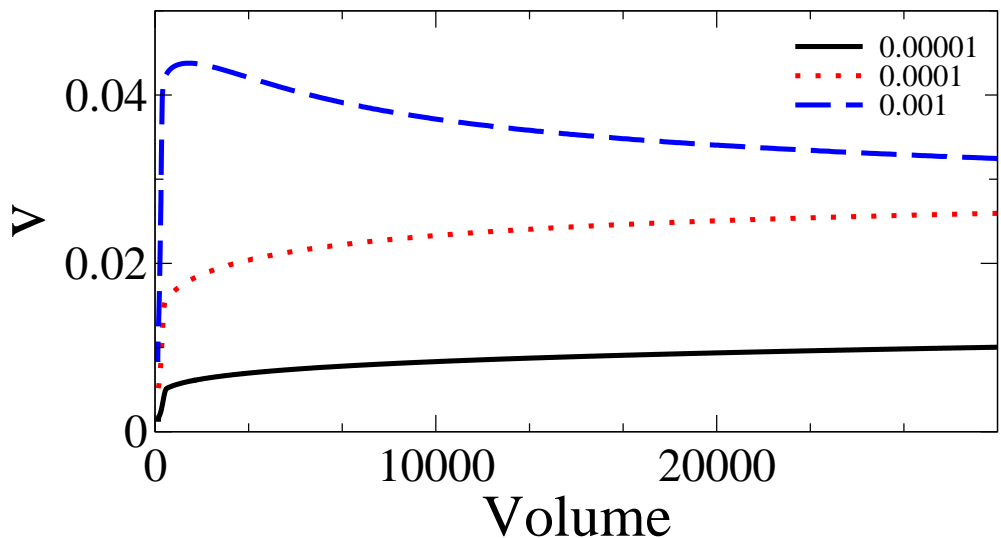}
\end{center}
\caption{The transition between (i) reaction-limited and 
(ii) saturation-limited regime is most clearly seen
in the dependence of velocity $v$ on drop size, i.e.\ droplet volume $V$. 
Curves are plotted for different reaction rates
$r$ (see legend). The remaining parameters are as in Fig.\,\ref{profiles}.
}
\label{length}
\end{figure}

Figs.\,\ref{vel}\,(b) and (c) give, respectively, 
the tangents of the measured advancing and receding dynamic contact 
angles ($\tan\theta_{a/r}$) and their differences 
$\Delta\tan\theta_{a/r}=\tan\theta_{a/r}-\tan\theta_e(\alpha_{a/r})$ from the
static contact angles as a function of the droplet's velocity in the
case $d=0$.
All contact angles given correspond to the mesoscopic contact angles
measured at the inflection points of the computed drop profiles. 
To determine the static contact angle we take the value of $\alpha$ at the inflection point
to calculate a space independent polar spreading coefficient 
in Eq.\,(\ref{sys1}).  
Fig.\,\ref{vel}\,(b) shows that the two dynamic contact angles
may differ quite strongly. 
However, a comparison of  Figs.\,\ref{vel}\,(b) and (c) shows that the difference between
the static and the dynamic contact angles is much smaller. 

Experimentally, an important measure is the dependence of droplet velocity 
on droplet length or volume \cite{DoOn95,LKL02}. 
We show in Fig.\,\ref{length} the dependence of $v$ on the droplet volume. 
For small reaction rates $r$ we find that the velocity increases with
drop length. 
The increase becomes more pronounced for larger $r$. 
This corresponds to the
reaction-limited regime {\em and} the experiments in \,\cite{DoOn95}. 
Upon further increase of $r$, 
the curve shows a velocity maximum at an intermediate drop size. 
For bigger drops, the velocity decreases monotonic similar to the
observations of  \cite{LKL02}.
Inspection of the corresponding profiles suggests that the regime of 
decreasing velocity is again the saturated regime.

To conclude, we have modelled chemically driven moving droplets
by a dynamical model consisting of coupled evolution equations for the film
thickness profile and the concentration of a surface coating that is produced
underneath the droplet. 
This model exhibits moving droplets with constant speed and shape in 
two- and three dimensions. 
A systematic analysis of the velocities of two-dimensional droplets
allows us to distinguish the reaction limited and saturated regimes, 
which are characterized by an increase resp.\ decrease of droplet
velocities with increasing droplet size or reaction rate. 
The behavior in the saturated regime indicates that the velocity of the droplets in
our model depend not only on the values of the chemical coating at 
the advancing and receding ends of the droplet, but are in addition
determined by the concentration profiles of $\alpha$ near the contact 
lines. 
The two regimes offer a natural explanation of the conflicting
experimental observation in \cite{DoOn95,LKL02}. 

Financial support by German-Israeli Foundation (GIF) is gratefully acknowledged.

%
%
%

%
%

\begin{thebibliography}{10}
\providecommand*{\bibinfo}[2]{#2}
\providecommand*{\eprint}[1]{#1}
\providecommand*{\url}[1]{#1}
\bibitem{Vela98}
\bibinfo{author}{M.~G. Velarde}, \bibinfo{journal}{Philos. Trans. R. Soc. Lond.
  Ser. A-Math. Phys. Eng. Sci.} \bibinfo{volume}{\textbf{356}},
  \bibinfo{pages}{829} (\bibinfo{date}{1998}).
\bibitem{Broc89}
\bibinfo{author}{F.~Brochard}, \bibinfo{journal}{Langmuir}
  \bibinfo{volume}{\textbf{5}}, \bibinfo{pages}{432} (\bibinfo{date}{1989}).
\bibitem{Raph88}
\bibinfo{author}{E.~Rapha{\"e}l}, \bibinfo{journal}{C. R. Acad. Sci. Ser. II}
  \bibinfo{volume}{\textbf{306}}, \bibinfo{pages}{751} (\bibinfo{date}{1988}).
\bibitem{ChWh92}
\bibinfo{author}{M.~K. Chaudhury} and \bibinfo{author}{G.~M. Whitesides},
  \bibinfo{journal}{Science} \bibinfo{volume}{\textbf{256}},
  \bibinfo{pages}{1539} (\bibinfo{date}{1992}).
\bibitem{RRV94}
\bibinfo{author}{A.~Y. Rednikov}, \bibinfo{author}{Y.~S. Ryazantsev}, and
  \bibinfo{author}{M.~G. Velarde}, \bibinfo{journal}{Phys. Fluids}
  \bibinfo{volume}{\textbf{6}}, \bibinfo{pages}{451} (\bibinfo{date}{1994}),
  and references therein.
\bibitem{MiMe97}
\bibinfo{author}{A.~S. Mikhailov}, and \bibinfo{author}{D. Meink\"ohn},
\bibinfo{pages}{334}, in \bibinfo{editor}{L. Schimansky-Geier} and 
\bibinfo{editor}{T. P\"oschel} (Eds.),  
\bibinfo{title}{\emph{Lecture Notes in Physics}}
\bibinfo{volume}{\textbf{484}}, (\bibinfo{publisher}{Springer}, Berlin,
\bibinfo{date}{1997}).
\bibitem{BBM94}
\bibinfo{author}{C.~D. Bain}, \bibinfo{author}{G.~D. Burnetthall}, and
  \bibinfo{author}{R.~R. Montgomerie}, \bibinfo{journal}{Nature}
  \bibinfo{volume}{\textbf{372}}, \bibinfo{pages}{414} (\bibinfo{date}{1994}).
\bibitem{DoOn95}
\bibinfo{author}{D.~F. Dos~Santos} and \bibinfo{author}{T.~Ondarcuhu},
  \bibinfo{journal}{Phys. Rev. Lett.} \bibinfo{volume}{\textbf{75}},
  \bibinfo{pages}{2972} (\bibinfo{date}{1995}).
\bibitem{LKL02}
\bibinfo{author}{S.~W. Lee}, \bibinfo{author}{D.~Y. Kwok}, and
  \bibinfo{author}{P.~E. Laibinis}, \bibinfo{journal}{Phys. Rev. E}
  \bibinfo{volume}{\textbf{65}}, \bibinfo{pages}{051602}
  (\bibinfo{date}{2002}).
\bibitem{ZWT98}
\bibinfo{author}{D.~W. Zheng}, \bibinfo{author}{W.~J. Wen}, and
  \bibinfo{author}{K.~N. Tu}, \bibinfo{journal}{Phys. Rev. E}
  \bibinfo{volume}{\textbf{57}}, \bibinfo{pages}{R3719} (\bibinfo{date}{1998}).
\bibitem{Haya02}
\bibinfo{author}{Y.~Hayashima} {\it et al.},
  \bibinfo{journal}{Phys. Chem. Chem. Phys.} \bibinfo{volume}{\textbf{4}},
  \bibinfo{pages}{1386} (\bibinfo{date}{2002}).
\bibitem{SBH00}
\bibinfo{author}{A.~K. Schmid}, \bibinfo{author}{N.~C. Bartelt}, and
  \bibinfo{author}{R.~Q. Hwang}, \bibinfo{journal}{Science}
  \bibinfo{volume}{\textbf{290}}, \bibinfo{pages}{1561} (\bibinfo{date}{2000}).
\bibitem{Brde95}
\bibinfo{author}{F.~Brochard-Wyart} and \bibinfo{author}{P.-G. de~Gennes},
  \bibinfo{journal}{C. R. Acad. Sci. Ser. II} \bibinfo{volume}{\textbf{321}},
  \bibinfo{pages}{285} (\bibinfo{date}{1995}).
\bibitem{ODB97}
\bibinfo{author}{A.~Oron}, \bibinfo{author}{S.~H. Davis}, and
  \bibinfo{author}{S.~G. Bankoff}, \bibinfo{journal}{Rev. Mod. Phys.}
  \bibinfo{volume}{\textbf{69}}, \bibinfo{pages}{931} (\bibinfo{date}{1997}).
\bibitem{Isra92}
\bibinfo{author}{J.~N. Israelachvili}, \bibinfo{title}{\emph{Intermolecular and
  Surface Forces}} (\bibinfo{publisher}{Academic Press}, London,
  \bibinfo{year}{1992}).
\bibitem{Shar93}
\bibinfo{author}{A.~Sharma}, \bibinfo{journal}{Langmuir}
  \bibinfo{volume}{\textbf{9}}, \bibinfo{pages}{861} (\bibinfo{date}{1993}).
\bibitem{TNPV02}
\bibinfo{author}{U.~Thiele} {\it et al.},
  \bibinfo{journal}{Colloid Surf. A} \bibinfo{volume}{\textbf{206}},
  \bibinfo{pages}{135} (\bibinfo{date}{2002}).
\bibitem{AUTO97}
\bibinfo{author}{E.~J. Doedel} {\it et al.},
  \bibinfo{title}{\emph{AUTO97: Continuation and bifurcation software for
  ordinary differential equations}} (\bibinfo{publisher}{Concordia University},
  Montreal, \bibinfo{year}{1997}).
%
\bibitem{boundary}
The stationary profiles are computed in a box of length $S$ and in a comoving
  frame $x - v t$. Boundary conditions are $h(S) = h(0)$ and $d \partial_x
  \alpha(z=S) + v \alpha(z=S) = 0$ resp.\ $ \partial_x \alpha(z=0) =0$
stemming from $\alpha(S) \to 0$ as $S \to \infty$ and the assumption of
  constant $\alpha$-profile behind the drop towards $- \infty$.
The streamlines in Fig.\,\ref{profiles} correspond to lines of constant
streamfunction $\psi(x,z)$ in the comoving frame given by
\begin{equation}
\psi(x,z)\,=\, \left( \frac{3z^2h}{2} - \frac{z^3}{2}\right)\,\partial_x
p(h,\alpha) + v z
\end{equation}
where the pressure $p(h,\alpha)$ corresponds to the expression in square
brackets in Eq.\,(\ref{sys1}).
%
\bibitem{note2}
The integration in time starts from
a liquid 'cylinder' with a small added noise using periodic boundary 
conditions. Initially, the $\alpha$ field is slightly shifted 
to fix the direction of movement. In the course of the
simulation it is set to its initial value far behind the droplet.
An explicite scheme is used on $96\times96$ and $128\times128$ grids 
and for drops moving in different directions with respect to the underlying 
grid.
%
\bibitem{KrMi94}
\bibinfo{author}{K.~Krischer} and \bibinfo{author}{A.~Mikhailov},
  \bibinfo{journal}{Phys. Rev. Lett.} \bibinfo{volume}{\textbf{73}},
  \bibinfo{pages}{3165} (\bibinfo{date}{1994}).
\bibitem{OBSP98}
\bibinfo{author}{M.~Or-Guil} {\it et al.}, \bibinfo{journal}{Phys. Rev. E}
  \bibinfo{volume}{\textbf{57}}, \bibinfo{pages}{6432} (\bibinfo{date}{1998}).
\bibitem{Boed03}
\bibinfo{author}{H.~U. B\"odeker} {\it et al.}, 
  \bibinfo{journal}{Phys. Rev. E} \bibinfo{volume}{\textbf{67}},
  \bibinfo{pages}{056220} (\bibinfo{date}{2003}).

\end{thebibliography}
\end{document}